\documentclass[prl,aps,twocolumn,showpacs,preprintnumbers,amsfonts]{revtex4}

\usepackage{dcolumn}% Align table columns on decimal point
\usepackage{latexsym}
\usepackage{graphics}
\usepackage{epsfig}
\newcommand{\bmat}{\left(\begin{array}}
\newcommand{\emat}{\end{array}\right)}
\newcommand{\beq}{\begin{equation}}
\newcommand{\eeq}{\end{equation}}
\newcommand{\drawsquare}[2]{\hbox{%
\rule{#2pt}{#1pt}\hskip-#2pt%  left vertical
\rule{#1pt}{#2pt}\hskip-#1pt%  lower horizontal
\rule[#1pt]{#1pt}{#2pt}}\rule[#1pt]{#2pt}{#2pt}\hskip-#2pt%  upper horizontal
\rule{#2pt}{#1pt}}% right vertical

\newcommand{\fund}{\raisebox{-.5pt}{\drawsquare{6.5}{0.4}}}%  fund
\newcommand{\Ysymm}{\raisebox{-.5pt}{\drawsquare{6.5}{0.4}}\hskip-0.4pt%
        \raisebox{-.5pt}{\drawsquare{6.5}{0.4}}}%  symmetric second rank
\newcommand{\Yasymm}{\raisebox{-3.5pt}{\drawsquare{6.5}{0.4}}\hskip-6.9pt%
        \raisebox{3pt}{\drawsquare{6.5}{0.4}}}%  antisymmetric second rank
\newcommand{\antifund}{\overline{\fund}}

\def\yzero{\smash{\hbox{$y\kern-4pt\raise1pt\hbox{${}^\circ$}$}}}

\def\-{\hphantom{-}}
\def\ov{\overline}
\def\s2{\frac{1}{\sqrt2}}

\def\beq{\begin{equation}}
\def\eeq{\end{equation}}
\def\beqa{\begin{eqnarray}}
\def\eeqa{\end{eqnarray}}

\def\IF{\relax{\rm I\kern-.18em F}}
\def\II{\relax{\rm I\kern-.18em I}}
\def\IP{\relax{\rm I\kern-.18em P}}

\def\Dsl{\,\raise.15ex\hbox{/}\mkern-13.5mu D} %can be subscripted

\def\IC{\bf C}
\def\IZ{\bf Z}
\def\IT{\bf T}
\def\z2z2{$\IC^3/(\IZ_2\times\IZ_2)$}

\def\s{\sigma}

\def\z{\zeta}

\def\bo{{\raise-.3ex\hbox{\large$\Box$}}}               % D'Alembertian
\def\face{{\raise.2ex\hbox{$\displaystyle \bigodot$}\mskip-2.2mu \llap {$\ddot
        \smile$}}}                                      % happy face
\def\leftrightarrowfill{$\mathsurround=0pt \mathord\leftarrow \mkern-6mu
        \cleaders\hbox{$\mkern-2mu \mathord- \mkern-2mu$}\hfill
        \mkern-6mu \mathord\rightarrow$}       % <--> double differential
\def\dvec#1{\vbox{\ialign{##\crcr
        \leftrightarrowfill\crcr\noalign{\kern-1pt\nointerlineskip}
        $\hfil\displaystyle{#1}\hfil$\crcr}}}           % <--> accent
\newcommand{\be}[3]{\begin{equation}  \label{#1#2#3}}
\newcommand{\bea}[3]{\begin{eqnarray}  \label{#1#2#3}}
\newcommand{\ee}{\end{equation}}
\newcommand{\eea}{\end{eqnarray}}
\newcommand{\ba}{\begin{array}}
\newcommand{\ea}{\end{array}}

\begin{document}

%%%%%%%%%%%%%%%%%%%%%%%%%%%%%%%%%%%%%%%%%%%%%%%%%%%%

\preprint{UPR-1088-T/rev, hep-th/0409032}

\title{Supersymmetric Standard Models, Flux Compactification and Moduli
Stabilization
}

\author{Mirjam Cveti{\v c} and Tao Liu}
\email{cvetic@cvetic.hep.upenn.edu}
\email{liutao@physics.upenn.edu}
\affiliation{%
Department of Physics and Astronomy, \\
University of Pennsylvania, Philadelphia, PA 19104-6396, USA}

\date{\today}

\begin{abstract}
Based on the T-dual constructions of supersymmetric intersecting D6-models on
$Z_2\times Z_2$ orientifolds, whose electroweak sector is parallel with the orientifold
planes with $Sp(2f)_L\times Sp(2f)_R$ gauge symmetry  (hep-th/0407178), we derive and
classify Standard Model-like vacua with RR and NSNS fluxes, which  stabilize toroidal
complex structure moduli and the dilaton. We find  consistent four-family ($f=4$)  and
two-family ($f=2$) models with  one- and two-units of  the quantized flux,
respectively. Such models typically possess
 additional gauge group factors with
negative  beta functions and may lead, via gaugino condensation, to stabilization
of toroidal K\"ahler moduli. These models have chiral exotics.
\end{abstract}


\maketitle

%%%%%%%%%%%%%%%%%%%%%%%%%%%%%%%%%%%%%%%%%%%%%%%%%%%%%%
{\bf Introduction\  } Constructions of explicit supersymmetric
Standard Models with
supergravity fluxes turned on is an outstanding technical problem
in string theory.  Supersymmetric semi-realistic constructions
with intersecting D6-branes  provide a beautiful, geometric
framework to engineer Standard Models with three families. The
first such  supersymmetric models were based on $Z_2\times Z_2$
orientifolds \cite{CSUI,CSUII} [Non-supersymmetric constructions
were given in \cite{LU,IB,IBII,LUII} (see also \cite{Sagnottietal}
and for earlier work \cite{Bachas,BDL}).] Turning on
supergravity fluxes will introduce a supergravity potential,
helping stabilize the compactification moduli fields by lifting
continuous moduli space of the string vacua in the effective
four-dimensional theory (see e.g. \cite{GVW}). However,
introducing the supergravity fluxes further restricts the
constructions, since such fluxes modify the global conditions on
the tadpole cancellations. The fluxes will typically
generate a back reaction on the original geometry of the internal
space, thus changing the nature of the internal space.

On the Type IIA side the supersymmetry conditions of flux
compactifications are less understood. Nevertheless recent work
\cite{BCI,BCII} revealed  the existence of  unique  flux vacua for
massive Type IIA string theory  with SU(3) structure, whose
geometry of the internal six-dimensional space is  nearly-K\"ahler
and four-dimensional space anti-deSitter. One such example is the
$ \ {SU(2)^3 \over SU(2)} \simeq S_3 \times S_3$  coset space that
has three supersymmetric three-cycles that   add up to zero in
homology \cite{Denefetal,BCII}. Therefore the total  charge of the
D6-branes wrapping such cycles is zero  and no introduction of
orientifold planes on such spaces is needed. Moreover, since the
three-cycles intersect  pair-wise,  the   massless chiral matter
appears at these intersections. This construction \cite{BCII}
therefore  provides an explicit example of  supersymmetric flux
compactifications  with intersecting D6-branes. Further progress
has  also been made in the constructions of N=1 supersymmetric
Type IIA flux vacua with SU(2)  structures \cite{BCII}, leading to
examples with  the   internal space comformally  Calabi-Yau.
However, explicit constructions of  models with intersecting probe
D6-branes for  such flux compactifications is  still awaiting
further study.

On the Type IIB side the intersecting D6-brane constructions
correspond to models with magnetized branes which have the role of
intersecting angle being played by the magnetic fluxes on the
branes. The dictionary for  consistency  and supersymmetry
conditions  between the two T-dual constructions is
straightforward, see e.g., \cite{CU}. The supersymmetric  Type IIB
flux compactifications are  also better understood. In particular,
examples of supersymmetric fluxes  and the internal space,
which is conformally Calabi-Yau, are well known
(see e.g. \cite{GKP} and
references therein). The prototype example is a self-dual
combination of the NSNS  $H_3$ and RR $F_3$ three-forms,
corresponding to the primitive (2,1) form on Calabi-Yau space.
Since the back-reaction of such flux configurations is mild, i.e.
the internal space remains conformal to Calabi-Yau, these  Type
IIB flux compactifications are especially suitable for adding the
probe magnetized D-branes in this background. However, the
quantization conditions on fluxes and the modified tadpole
conditions constrain the possible D-brane configurations severely.
In Refs.   \cite{CU,BLT} techniques  for consistent chiral flux
compactifications on orbifolds were developed, however no explicit
supersymmetric chiral Standard Model flux compactifications  were
obtained. Most recently  in \cite{MS} an example of a one-family
Standard Model with the supersymmetric (three units of quantized)
flux was constructed. This construction also yields  three- and
two-family models with one- and two-units of the quantized flux,
thus providing the first examples of semi-realistic Standard Model
flux compactifications. The D-brane sector construction is T-dual
to models of intersecting D6-branes on the $Z_2\times Z_2$
orientifold with the $Sp(2)_L\times Sp(2)_R$ symmetry in the
electroweak sector \cite{CIM,CLLL}. [Without fluxes, the first
models of that type were toroidal models  with intersecting
D6-branes \cite{CIM} where the RR tadpoles were not explicitly
cancelled. The $Z_2\times Z_2$ orientifold construction in
\cite{CLLL} cancelled the RR-tadpoles by introducing an additional
stack of branes with unitary symmetry. However, the original
additional stack had $U(1)$ gauge symmetry, which introduced the
discrete global anomaly \cite{WITT}, as pointed out in
\cite{MSII}. The anomaly is due to an odd-number of intersections
of $U(1)$ branes with $Sp(2N)$ branes, thus resulting in an
odd-number of chiral superfields in the fundamental representation
of $Sp(2N)$; the non-anomalous models can be easily
constructed by introducing analogous $U(2)$ branes instead, as in the
revised version of \cite{CLLL}.]

The aim of this paper is to  study the construction of
supersymmetric Standard Models with fluxes turned on. We shall
employ the specific $Z_2\times Z_2$ orientifold constructions that
are T-dual to intersecting D6-brane constructions whose
electroweak sector arises from D6-branes originally residing on
top of the orientifold planes. These constructions (without
fluxes) were discussed in detail in \cite{CLLL}. The first class
is based on one-family $U(4)\times Sp(2f)_L\times Sp(2f)_R$ gauge
symmetry (in the observable sector),  which yields  the f-family
Standard Model gauge symmetry (with additional $U(1)$'s) by
employing the brane-splitting mechanism. [For $f=4$ and $f=3$ such
constructions (without fluxes) yield \cite{CLLL}  anomaly free
models without chiral exotics.] The  second class is  based on
constructions with  $Sp(2)_L\times Sp(2)_R$ as the starting
electroweak symmetry \cite{CIM,CLLL,MS}. Within this framework we
first present two classes of constructions with supersymmetric
fluxes (three-units of   quantized flux) which, however, suffer
from the discrete global  anomaly \cite{WITT}.

We further classify anomally-free  flux compactifications based on the brane
constructions  with $Sp(2f)_L\times  Sp(2f)_R$ electroweak sector: we find  three- and
four-family models ($f=3,4$) with one-unit of quantized flux, as well as  one- and
two-family  models ($f=1,2$) with two-units of quantized  flux. [Note however, that for
$f=3$ the breaking pattern of  $Sp(6)_L\times Sp(6)_R$ down to $Sp(2)_L\times Sp(2)_R$
breaks supersymmetry \cite{CLLL}.] For all constructed models, the turned on fluxes
will fix the toroidal complex structure moduli \cite{KST}.  In addition, these
models  typically possess  a ``hidden sector'' group factor with  a
 negative beta function; if such a gauge factor is  associated with  D7-branes, the
 non-perturbative infrared dynamics
on these  D7-brane  may lead to gaugino condensation and stabilization of a toroidal K\"ahler modulus.
 The additional toroidal K\"ahler moduli  may be fixed due to the
supersymmetry constraints. [Note that the open string moduli and the K\"ahler moduli
can form combined  flat directions, allowing for brane recombinations (see
\cite{CSUII}). However, due to the flux back-reaction it is expected that the open
string moduli become massive \cite{CU}; in this case the supersymmetry conditions fix
K\"ahler moduli.]

{\bf Model\  } The generators $\theta$ and $\omega$ for the orbifold
group $Z_{2}
\times Z_{2}$, act on the complex coordinates of $T^6$ as $\theta:
(z_1,z_2,z_3) \to (-z_1,-z_2,z_3)$ and $\omega: (z_1,z_2,z_3) \to
(z_1,-z_2,-z_3)$ and  the projection $\Omega R$, where $\Omega$ is
the world-sheet parity projection and R (acting on Type IIA  as
the holomorphic $Z_2$ involution): $(z_1,z_2,z_3) \to
(-z_1,-z_2,-z_3)$. The actions of $\Omega R$, $\Omega R\omega$,
$\Omega\theta\omega$ and $\Omega R\theta$ introduce, on the Type
IIB side (i.e. after T-dualizing the Re$(z_1)$, Re$(z_2)$, and
Re$(z_3 )$ toroidal directions), 64 $O3$-planes and 4
$O7_i$-planes, each of them sitting on $Z_2$ fixed points of the
$i^{th}$ $T^2$ and wrapping the other two.

To compensate for  the negative RR charges from these O-planes, we
need to introduce D(3+2n)-branes in our models which are filling
up $D=4$ Minkowski space and wrapping 2n-cycles on the compact
manifold. These D(3+2n)-branes can be magnetized (The detailed
discussion for  toroidal/orbifold  compactifications with
magnetized branes  is given in \cite{CU}.) Concretely, for one
stack of $N_a$ D-branes wrapped $m_a^i$ times on the $i^{th}$
2-torus $T^2_i$, we turn on $n_a^i$ units of magnetic fluxes $F_a$
for  the  $U(1)_a$ gauge factor (associated with the D-brane
center of mass motion)  on each $T^2_i$, such that
\begin{eqnarray}
m_a^i \, \frac 1{2\pi}\, \int_{\IT^2_{\,i}} F_a^i \, = \, n_a^i\, .
\label{monopole}
\end{eqnarray}
Hence the topological information of this stack of D-branes is
encoded in  $N_a$-number of D-branes  and the  co-prime number
pairs $(n_a^i,m_a^i)$.

The chiral massless spectrum can be
generated from the ``intersection" of two stacks of D-branes a and
b denoted by the intersection number
\begin{eqnarray}
I_{ab}=\prod_{i=1}^3(n_a^im_b^i-n_b^im_a^i) \label{intersections}
\end{eqnarray}
The point for the spectra is that they should be invariant under
the full orientifold symmetry group. A detailed discussion of Type
IIA has been given in \cite{CSUI,CSUII}, which can be easily
interpreted into type IIB (see Table \ref{spectrum}).

\begin{table}
[htb]\footnotesize
\renewcommand{\arraystretch}{1.25}
\caption{General spectrum on magnetized D-branes in type IIB
$T^6/(Z_2\times Z_2)$ orientifold. The representations in the
table refer to $U(N_a/2)$, the resulting gauge symmetry due to
$Z_2\times Z_2$ orbifold projection. In our convention, positive
intersection numbers implies left-hand chiral supermultiplets.}
\begin{center}
\begin{tabular}{|c|c|} \hline {\bf Sector} & {\bf Representation} \\
\hline\hline
$aa$   & $U(N_a/2)$ vector multiplet  \\
       & 3 adjoint chiral multiplets  \\
\hline
$ab+ba$   & $I_{ab}$ $(\fund_a,\antifund_b)$ fermions   \\
\hline
$ab'+b'a$ & $I_{ab'}$ $(\fund_a,\fund_b)$ fermions \\
\hline $aa'+a'a$ &$\frac 12 (I_{aa'} - \frac 12 I_{a,O})\;\;
\Ysymm\;\;$ fermions \\
          & $\frac 12 (I_{aa'} + \frac 12 I_{a,O}) \;\;
\Yasymm\;\;$ fermions \\
\hline
\end{tabular}
\end{center}
\label{spectrum}
\end{table}

D(3+2n)-branes can induce D-brane charges of lower odd dimension
due to world-volume couplings. Explicitly, for one stack of $N_a$
D-branes with wrapping number $(n^i_a,m^i_a)$, it carries D3-,
D5-, D7 and D9-brane RR charges
\begin{eqnarray}
Q3_a=N_an_a^1n_a^2n_a^3,& (Q5_i)_a=N_am_a^in_a^jn_a^k\, , \nonumber\\
(Q7_i)_a=N_an_a^im_a^jm_a^k,&Q9_a=N_am_a^1m_a^2m_a^3\, ,
\end{eqnarray}
where $i\neq j\neq k$ and a permutation is implied for $(Q5_i)_a$
and $(Q7_i)_a$. Besides the D-brane and O-plane, fluxes can also
contribute RR tadpole. A consistent string model requires that the
RR sources satisfy the Gauss law in the compact space, $i.e$, RR
tadpole must be cancelled. For D3- and D7-branes, thus we have
\begin{eqnarray}
-N^{(0)}+\sum_a Q3_a-\frac{1}{2}N_{flux}=-16 \nonumber \\
-N^{(i)}+\sum_a (Q7_i)_a=-16, i\neq j\neq k  \, , \label{RR tadpole}
\end{eqnarray}
where $N^{(0)}$ and $N^{(i)}$ with $i=1, 2, 3$ and 4 respectively
denotes the number of filler branes, $i.e.$, D-branes which wrap
along O3- and $O7_i$-planes and only contribute one of the four
kinds of D3- and D7-brane charges. As for D5- and D9-brane RR
tadpoles, their cancellation is automatic since D-branes and their
$\Omega R$ images carry the same number charges with different
signs.

Four-dimensional $N=1$ supersymmetric vacua from flux
compactification require that 1/4 supercharges from
ten-dimensional Type I T-dual be preserved in both open string and
closed string sectors. On the type IIB orientifold, supersymmetric flux
solutions have been discussed in \cite{GKP,KST}.  The
configuration has RR $F_3$  and NSNS $H_3$  three-form flux
turned. The supersymmetry conditions imply  that the  three-form
$G_3=F_3-\tau H_3$ is a primitive , self-dual (2,1) form. Here
$\tau=a+i/g_s$ is  type IIB axion-dilaton coupling. The  $G_3$
flux will contribute a D3-brane RR charge
\begin{eqnarray}
N_{\rm flux}\, =\, \frac{1}{(4\pi^2\alpha')^2} \,
\frac{i}{2\tau_I}\, \int_{X_6} \, G_3\wedge {\ov G}_3\,,
\end{eqnarray}
where $\tau_I$  is the imaginary part of the complex coupling $\tau$.
Dirac quantization conditions of $F_3$ and $H_3$ on
$T_6/(Z_2\times Z_2)$ orientifold require that $N_{flux}$ is
 a multiple of 64 and and  BPS-like  self-duality condition: $*_6G_3=iG_3$ ensures that its contribution to
the  RR  charge is positive. We shall employ the following  specific solution \cite{KST,CU}:
\begin{eqnarray}
G_3\, =\textstyle{\frac{8}{\sqrt{3}}}\, e^{-\pi i/6}\, (\, d{\ov
z}_1dz_2dz_3\, + \, dz_1d{\ov z}_2dz_3\, +\, dz_1dz_2d{\ov
z}_3\,), \label{G3}
\end{eqnarray}
where the additional factor 4 is due to the $Z_2\times Z_2$
orbifold symmetry. The flux stabilizes the complex structure  toroidal moduli at values
\begin{eqnarray}
\tau_1=\tau_2=\tau_3=\tau=e^{2\pi i/3},
\end{eqnarray}
leading to the   RR tadpole  contribution in
Eq(\ref{RR tadpole}): $N_{\rm flux}=192$.

For magnetized  D-branes with world-volume magnetic field
$F^i=\frac{n^i}{m^i\chi^i}$ in the open string sector, the
four-dimensional $N=1$ SUSY can be preserved by the orientation
projection if and only if its rotation angles with respect to the
orientifold-plane are an element of $SU(3)$ rotation, thus
implying: $\theta_1+\theta_2+\theta_3=0 $ mod $2\pi$. Here
$\chi^i=R^i_1R^i_2$, the   area of the $i^{th}$ $T^2$ in $\alpha'$
units, is the K\" ahler modulus and
$\theta_i=arctan[(F^i)^{-1}]$ is the "angle" between the D-brane
and the O-plane in the $i^{th}$ $T^2$.

{\bf Frameworks\  }
We choose to construct the Standard-like Models as descendants of
the Pati-Salam model based on $SU(4)_C\times SU(2)_L\times
SU(2)_R$.  The hypercharge is thus:
\begin{eqnarray}
Q_Y=Q_{I_{3R}}+{{Q_{B-L}}\over{2}}\, ,
\end{eqnarray}
where  the non-anomalous $U(1)_{B-L}$ is obtained from the
splitting of the $U(4)_C$ branes $\to U(3)_C\times U(1)_{B-L}$.
Similarly, anomaly-free $U(1)_{I_{3R}}$ is part of the non-Abelian
part of $U(2)_R$ or $Sp(2)_R$ gauge symmetry.
Within this
framework we found two classes of constructions  which may yield
Standard models with fluxes turned on \cite{foot}:
%Both classes of these
%constructions (without fluxes) have been explored within
%intersecting D6-branes scenario in \cite{CLLL}:

{ (i)} The starting  symmetry is one-family $U(4)\times
Sp(2f)_L\times Sp(2f)_R$,  ($f=4$) which can be broken down to the
four-family $U(4)\times U(2)_L\times U(2)_R$ by  parallel splitting
the D-branes, originally positioned on the O-planes, in some
two-tori directions.  [Both the string theory and field theory
aspects of the brane splitting in this framework is discussed in
detail in \cite{CLLL}.] In the field theory picture
(``Higgsing''), the four-families ($f=4$) are obtained as a
decomposition
of the original chiral supermultiplets $(4,8,1)$ and $(4,1,8)$
into four copies of $(4,2,1)$ and $(4,1,2)$.
 The Higgsing,  as discussed in \cite{CLLL}, preserves  D- and
F-flatness and thus symmetry breaking   can  take place at the
string scale, generating a 4D $N=1$ supersymmetric four-family
Pati-Salam model. Note furthermore, that in this case the $U(1)_L$
and $U(1)_R$ are not anomalous since they arise from the
non-Abelian $Sp$ symmetry. One expects that at least the gauge
boson of $U(1)_L$ will have a mass at the electroweak scale.
%There
%is additional $U(1)$ factor in the ``hidden sector'' under which
%Standard Model chiral exotic particles are charged;  they may
%become massive at the $U(1)$ breaking scale.
[This analysis can
also  be applied to two- ($f=2$) and  three- ($f=3$) family
examples.  Note however, that while for $f=2$ the Higgsing
$Sp(4)_L\times Sp(4)_R\to Sp(2)_L\times Sp(2)_R$  is
supersymmetric, for  $f=3$    the  Higgsing $Sp(6)_L\times Sp(6)_R
\to Sp(2)_L\times Sp(2)_R$ breaks supersymmetry \cite{CLLL}.]

{(ii)}  The  starting symmetry  is Pati-Salam-like $U(4)_C\times
Sp(2)_L\times Sp(2)_R$.
%, discussed in \cite{CLLL}. [The toroidal
%orientifold  examples that did not explicitly cancel RR-tadpoles
%were discussed in \cite{CIM}.)
Recently,  in \cite{MS} the constructions of this type were explored to
obtain the first examples of semi-realistic  Standard Model flux
compactifications.

The gauge symmetry in the observable sector is $SU(4)\times
SU(2)_L\times SU(2)_R$, with the subsequent symmetry breaking
chain \cite{CLLL}:
\begin{eqnarray}
&& SU(4)\times SU(2)_L \times SU(2)_R \nonumber\\
\rightarrow && SU(3)_C\times SU(2)_L \times SU(2)_R \times
U(1)_{B-L} \nonumber\\
 \rightarrow && SU(3)_C\times SU(2)_L\times U(1)_Y~.~\,
\end{eqnarray}
Here the first step can be achieved by splitting the $a^{th}$
stack of D-branes at string scale and the second one by giving
VEVs to the scalar component of right-handed neutrino superfield
at TeV scale. Therefore at the electroweak scale one would only
have the minimal supersymmetric Standard Model content.
\begin{table}
[htb] \footnotesize
\renewcommand{\arraystretch}{1.0}
\caption{D-brane configurations and intersection numbers for the
four-family supersymmetric Standard-like Models with three-units of
quantized  flux. Here, $r=3,4$,
$\chi_i$ is the K\" ahler modulus for the $i^{th}$ two-torus, and
$\beta_j^g$ is the beta function for the $Sp$ group from the
$j^{th}$ stack of branes. These models have discrete global
anomaly.} \label{Sp(8) X Sp(8)}
\begin{center}
\begin{tabular}{|c||c|c||c|c|c|c|c|c|c|c|}
\hline
    \multicolumn{11}{|c|}{$[U(4)_C\times Sp(8)_L\times Sp(8)_R]_{o} \times [U(1)\times Sp(32r-88)\times Sp(8-2r)]_{h}$}\\
\hline \hline \rm{j} & $N$ & $(n^1,m^1)(n^2,m^2)
(n^3,m^3)$ & $n_{\Ysymm}$& $n_{\Yasymm}$ & $b$  & $c$ & $d$ & $d'$ & 1 & 2  \\
\hline \hline
    $a$&  8& $(1,0)(1,1)(1,-1)$ & 0 & 0  &  1 & -1 & 15 & -15 & 0 & 0 \\
    $b$&  8& $(0,1)(1,0)(0,-1)$ & 0 & 0  &  - & 0 & 4r & -4r & 0 & 0 \\
    $c$&  8& $(0,1)(0,-1)(1,0)$ & 0 & 0  &  - & - & 4r & -4r & 0 & 0 \\
\hline \hline
    $d$&  2& $(-r,-1)(4,1)(4,1)$ & 72r & 56r  &  - & - & - & - & 1 & -16 \\
\hline
    1&   32r-88& $(1,0)(1,0)(1,0)$  & \multicolumn{8}{c|} {$\chi_1=8r\chi_2/(\chi_2^2-16), \chi_2=\chi_3$}\\
    2&   8-2r& $(1,0)(0,-1)(0,1)$ & \multicolumn{8}{c|} {$\beta^g_1=-11/2$}\\
\hline
\end{tabular}
\end{center}
\end{table}
\begin{table*}
[htb] \footnotesize
\renewcommand{\arraystretch}{1.0}
\caption{D-brane configurations and intersection numbers for three-
and four-family supersymmetric Standard-like Models with three units
$n_f=3$ of the quantized flux. $f$ is the family
number, and $2\leq r\leq 4$ for $f=3$ and $3\leq
r\leq 4$ for $f=4$. $\chi_i$ is the K\" ahler
modulus for the $i^{th}$ two-torus, and $\beta_j^g$ is the beta
function for the $Sp$ group from the $j^{th}$ stack of branes.
These models have discrete global anomaly.} \label{Sp(2) X Sp(2)}
\begin{center}
\begin{tabular}{|c||c|c||c|c|c|c|c|c|c|c|}
\hline
    \multicolumn{11}{|c|}{$[U(4)_C\times Sp(2)_L\times
    Sp(2)_R]_{o} \times [U(1)\times Sp(98r-8f^2-80)\times Sp(8-2r)]_{h}$}\\
\hline \hline \rm{j} & $N$ & $(n^1,m^1)(n^2,m^2)
(n^3,m^3)$ & $n_{\Ysymm}$& $n_{\Yasymm}$ & $b$  & $c$ & $d$ & $d'$ & 1 & 2  \\
\hline \hline
    $a$&  8& $(1,0)(f,1)(f,-1)$ & 0 & 0  &  f & -f & $49-f^2$ & $f^2-49$ & 0 & 0 \\
    $b$&  2& $(0,1)(1,0)(0,-1)$ & 0 & 0  &  - & 0 & 7r & -7r & 0 & 0 \\
    $c$&  2& $(0,1)(0,-1)(1,0)$ & 0 & 0  &  - & - & 7r & -7r & 0 & 0 \\
\hline \hline
    $d$&  2& $(-r,-1)(7,1)(7,1)$ & 210r+33 & 182r-33  &  - & - & - & - & 1 & -49 \\
\hline
    1&   $98r-8f^2-80$ & $(1,0)(1,0)(1,0)$  & \multicolumn{8}{c|} {$\chi_1=14r\chi_2/(\chi_2^2-49), \chi_2=\chi_3$}\\
    2&   $8-2r$ & $(1,0)(0,-1)(0,1)$ & \multicolumn{8}{c|} {$\beta^g_1=-11/2$}\\
\hline
\end{tabular}
\end{center}
\end{table*}

Within the above frameworks  classes four- and
three-family supersymmetric Standard Models  are presented in
Table \ref{Sp(8) X Sp(8)} and Table \ref{Sp(2) X Sp(2)}, $i.e.$, those
models have three-units of  quantized flux.  However,  the
branes with the  $U(1)$ symmetry have an odd number of intersections with the
$Sp$ branes, thus introducing an odd number of fundamental
representations of $Sp$, {\it i.e.} the models suffer from the
discrete global anomaly \cite{WITT}. The conditions for the
absence of discrete global anomalies can be derived from the
K-theory \cite{UR} and have been given for the $Z_2\times Z_2$
orientifolds in \cite{MSII}:
\begin{eqnarray}
\sum N_am_a^1m_a^2m_a^3=4Z,& \sum N_an_a^im_a^jm_a^k=4Z\, , \nonumber\\
\sum N_am_a^in_a^jn_a^k=4Z,& \sum N_an_a^1n_a^2n_a^3=4Z\, ,
\end{eqnarray}
where $Z$ is an integer. These conditions turn out to be extremely
constraining and  within the above framework no anomaly-free four- or
three-family models with   three units of  quantized flux
exist. Thus,  explicit constructions of  semi-realistic
models with supersymmetric fluxes  remain elusive.

{\bf Classification \ } We now turn to the
classification of the
supersymmetric D-brane constructions with $Sp(2f)_L\times
Sp(2f)_R$ electroweak sector (framework (i)) with the maximal
number of  flux units turned on [Within framework (ii) a related
analysis was done in \cite{MSII}.]
In the following $n_f=N_{flux}/64$, and
$(n_f)_{max}$ is the largest  allowed unit of the quantized flux:

{(1)} Four-families ($f=4$): $(n_f)_{max}=1$. We display a
representative example in  Table \ref{Sp(8) X Sp(8) non-SUSY1}
with the $U(2)$ symmetry in the hidden sector.  We also found
another model   with hidden sector symmetry $U(1)\times U(1)$
and respective brane configurations $(-1,-1)(3,1)(2,1)$ and
$(-3,-1)(1,1)(2,1) $.
% All possible models
%are generated within the framework (i)
%and in the frame
[Within  framework (ii) there are no models with $n_f\ne 0$.]

{(2)} Three-families ($f=3$): $(n_f)_{max}=1$. The solution
possesses the hidden sector gauge symmetry  with  unitary gauge
  sector  $U(1)\times U(1)$  and  the respective brane configurations
$(-2,-1)(2,1)(3,1)$ and $(-2,-1)(3,1)(2,1)$. However, since the
Higgsing of $Sp(6)_L\times Sp(6)_R \to Sp(2)_L\times Sp(2)_R$
breaks supersymmetry \cite{CLLL} these D-brane models  are not
supersymmetric. [Within  framework  (ii), the solution found in
\cite{MS} is the only possible one.]

{(3)} Two-families ($f=2$): $(n_f)_{max}=2$.  In  Table \ref{Sp(8)
X Sp(8) non-SUSY1}   we present a model with the $U(2)$ symmetry
in the hidden sector. We also found models with $U(1)\times U(1)$
in the hidden sector. [Witnin  framework (ii)  the only  example
is given in \cite{MS}.]

{(4)} One-family ($f=1$): $(n_f)_{max}=3$. In this case frameworks
(i) and  (ii) are  equivalent.  The only example with
 $(n_f)_{max}=3$ was obtained in \cite{MS}.  For the
case with one tilted two-torus, there are  models with $n_f=2$ and
$U(2)$  or $U(1)\times U(1)$ gauge symmetry in the hidden sector.

\begin{table*}
[htb] \footnotesize
\renewcommand{\arraystretch}{1.0}
\caption{D-brane configurations and intersection numbers for the
consistent $f$-family Standard-like Models with $n_f$-units of
quantized flux. $\chi_i$ is the K\" ahler modulus for the
$i^{th}$ two-torus, $\beta_j^g$ is the beta function for the $Sp$
group from the $j^{th}$ stack of branes.  The allowed models have
$f=2,4$  with $(n_f)_{max}=2,1$,
respectively.} \label{Sp(8) X Sp(8) non-SUSY1}
\begin{center}
\begin{tabular}{|c||c|c||c|c|c|c|c|c|c|}
\hline
    \multicolumn{10}{|c|}{$[U(4)_C\times Sp(2f)_L\times Sp(2f)_R]_{o} \times [U(2)\times Sp(8(4-\frac{f}{2})^2+16-32n_f)]_{h}$}\\
\hline \hline \rm{j} & $N$ & $(n^1,m^1)(n^2,m^2)
(n^3,m^3)$ & $n_{\Ysymm}$& $n_{\Yasymm}$ & $b$  & $c$ & $d$ & $d'$ & 1   \\
\hline \hline
    $a$&  8& $(1,0)(1,1)(1,-1)$ & 0 & 0  &  1 & -1 & $(4-\frac{f}{2})^2-1$ & $-(4-\frac{f}{2})^2+1$ & 0  \\
    $b$&  8& $(0,1)(1,0)(0,-1)$ & 0 & 0  &  - & 0 & $2(4-\frac{f}{2})$ & $-2(4-\frac{f}{2})$ & 0  \\
    $c$&  8& $(0,1)(0,-1)(1,0)$ & 0 & 0  &  - & - & $2(4-\frac{f}{2})$ & $-2(4-\frac{f}{2})$ & 0  \\
\hline \hline
    $d$&  4& $(-2,-1)(4-\frac{f}{2},1)(4-\frac{f}{2},1)$ & \multicolumn{7}{c|} {$\chi_1=(16-2f)\chi_3/(\chi_3^2-(4-\frac{f}{2})^2)$} \\
    1&   $8(4-\frac{f}{2})^2+16-32n_f$& $(1,0)(1,0)(1,0)$ & \multicolumn{7}{c|} {$\chi_2=\chi_3$, $\beta^g_1=-5$} \\
\hline
\end{tabular}
\end{center}
\end{table*}

%As for another possible frame with the electroweak sector given by
%the standard Pati-Salam gauge structure $U(2)_L\times U(2)_R$,
%which has been explored in the background of vanishing flux in
%\cite{CLL}, the situation becomes worse due to the relatively
%large RR charges carried by U(N) branes.

{\bf Moduli Stabilization\  }  These models   typically contain a
``hidden sector'', associated with D-branes,
 parallel with the O-planes (``filler'' branes),  whose
  gauge group factors have    negative beta functions.
 If such a gauge gauge factor were associated with a stack of D7$_i$-branes,
 the  infrared non-perturbative
gauge dynamics could generate a   superpotential
of the form (see, e.g., \cite{CLW}):
\begin{eqnarray}
W_{eff} = \frac{\beta_i \Lambda^3}{32 e \pi^2}
 \exp \left(\frac{8\pi^2}{\beta_i}
f_W(T_i)\right)+W_o\, ,
\end{eqnarray}
where  the gauge function $f_W(T_i) =
n^1_im^2_jm^3_k\,  T_i$  ($i\ne j \ne k\ne i$),  $T_i$ is the toroidal K\"ahler
modulus for the i-th two-torus,  $\beta_i$  is the beta function,
  $\Lambda$ is  the cutoff string scale
 and $W_o$ is a contribution from the fluxes \cite{KKLT}  which fixes the dilaton
and complex structure moduli.  [Due to the flux back-reaction  the open string sector
moduli are expected to get a mass \cite{CU}; in this case the supersymmetry conditions
fix the rest of toroidal K\" ahler moduli (say,  ${\rm Re} T_{j,k}\sim  \chi_{j,k}$ in
terms of $T_i$).] Models desplayed in Table \ref{Sp(8) X Sp(8) non-SUSY1}  possess only
the hidden sector  D3-branes with confining gauge symmetry. However, a generalization
\cite{Cvetic:2005bn} of the present constructions, does provide Standard-like  flux
models with such confining D7$_i$-branes. 
Note,  that for   supersymmetric flux ($n_f=3$) $W_o=0$, and for the 
non-supersymmetric ones ($n_f=1,2$)
$W_o\neq 0$. In the latter case the resulting $W_{eff}$, along with the K\"ahler
potential for the volume modulus, determine the supergravity potential whose ground
state would fix also the remaining toroidal K\"ahler modulus \cite{KKLT}. However,  for
the models under consideration $W_o={\cal O}(M_{P}^3)$ and  thus the volume modulus
$f_W={\cal O}(1)$. A mechanism to turn on a small value for $W_o$  remains an open
problem.

\begin{acknowledgments}
We thank Tianjun Li and Gary Shiu for discussions, and  are
especially grateful to Paul Langacker for many discussions and
Angel Uranga for useful communications. Research is supported in
part by DOE grant DOE-EY-76-02-3071 (M.C.,T.L.), NSF grant
INT02-03585 (M.C.) and Fay R. and Eugene L. Langberg endowed Chair
(M.C.).
\end{acknowledgments}


\begin{thebibliography}{}

\bibitem{CSUI}
M.~Cveti\v c, G.~Shiu and A.~M.~Uranga, Phys.\ Rev.\ Lett.\  {\bf
87}, 201801 (2001), hep-th/0107143.
%%CITATION = HEP-TH 0107143;%%


\bibitem{CSUII}
M.~Cveti\v c, G.~Shiu and A.~M.~Uranga, Nucl.\ Phys.\ {\bf B615},
3 (2001), hep-th/0107166.
%%CITATION = HEP-TH 0107166;%%


\bibitem{LU}
R.~Blumenhagen, L.~G\"orlich, B.~K\"ors and D.~L\"ust, JHEP {\bf
0010} (2000) 006,hep-th/0007024.
%CITATION = HEP-TH 0007024;%%

\bibitem{IB}
G.~Aldazabal, S.~Franco, L.~E.~Ib\'a\~nez, R.~Rabad\'an and
A.~M.~Uranga, JHEP {\bf 0102}, 047 (2001), hep-ph/0011132.
%%CITATION = HEP-PH 0011132;%%


\bibitem{IBII}
G.~Aldazabal, S.~Franco, L.~E.~Ib\'a\~nez, R.~Rabad\'an and
A.~M.~Uranga, J.\ Math.\ Phys.\  {\bf 42}, 3103 (2001), hep-th/0011073.
%%CITATION = HEP-TH 0011073;%%

\bibitem{LUII}
R.~Blumenhagen, B.~K\"ors and D.~L\"ust, JHEP {\bf 0102} (2001)
030,
hep-th/0012156.
%%CITATION = HEP-TH 0012156;%%

\bibitem{Sagnottietal}
C.~Angelantonj, I.~Antoniadis, E.~Dudas and A.~Sagnotti,
%``Type-I strings on magnetised orbifolds and brane transmutation,''
Phys.\ Lett.\ B {\bf 489}, 223 (2000),
hep-th/0007090.
%%CITATION = HEP-TH 0007090;%%


\bibitem{Bachas}
C.~Bachas,
%``A Way to break supersymmetry,''
hep-th/9503030.
%%CITATION = HEP-TH 9503030;%%

\bibitem{BDL}
M.~Berkooz, M.R.~Douglas and R.G.~Leigh,
%``Branes intersecting at angles,''
Nucl.\ Phys.\ B {\bf 480}, 265 (1996),
hep-th/960613.
%%CITATION = HEP-TH 9606139;%%

\bibitem{GVW}
S.~Gukov, C.~Vafa and E.~Witten,
%``CFT's from Calabi-Yau four-folds,''
Nucl.\ Phys.\ B {\bf 584}, 69 (2000)
[Erratum-ibid.\ B {\bf 608}, 477 (2001)],
hep-th/9906070.
%%CITATION = HEP-TH 9906070;%%


\bibitem{BCI}
K.~Behrndt and M.~Cveti\v c,
%``General N = 1 supersymmetric flux vacua of (massive) type IIA string
%theory,''
hep-th/0403049.
%%CITATION = HEP-TH 0403049;%%



\bibitem{BCII}
K.~Behrndt and M.~Cveti\v c,
%``General N = 1 supersymmetric fluxes in massive type IIA string theory,''
hep-th/0407263.
%%CITATION = HEP-TH 0407263;%%

\bibitem{Denefetal}
B.~S.~Acharya, F.~Denef, C.~Hofman and N.~Lambert,
%``Freund-Rubin revisited,''
hep-th/0308046.
%%CITATION = HEP-TH 0308046;%%

\bibitem{CU}
J.F.G. Cascales and A.M. Uranga,
%``Chiral 4d N = 1 string vacua with D-branes and NSNS and RR fluxes,''
JHEP {\bf 0305}, 011 (2003), hep-th/0303024.
%%CITATION = HEP-TH 0303024;%%


\bibitem{GKP}
S.B.~Giddings, S.~Kachru and J.~Polchinski,
%``Hierarchies from fluxes in string compactifications,''
Phys.\ Rev.\ D {\bf 66}, 106006 (2002), hep-th/0105097.
%%CITATION = HEP-TH 0105097;%%


\bibitem{BLT}
R.~Blumenhagen, D.~L\"ust and T.R. Taylor,
%``Moduli stabilization in chiral type IIB orientifold models with fluxes,''
Nucl.\ Phys.\ B {\bf 663}, 319 (2003),
hep-th/0303016.
%%CITATION = HEP-TH 0303016;%%

\bibitem{MS}
F.~Marchesano and G.~Shiu,
%``MSSM vacua from flux compactifications,''
hep-th/0408059.
%%CITATION = HEP-TH 0408059;%%

\bibitem{CIM}
D.~Cremades, L.E.~Ib\'a\~nez and F.~Marchesano,
%``Yukawa couplings in intersecting D-brane models,''
JHEP {\bf 0307}, 038 (2003), hep-th/0302105.
%%CITATION = HEP-TH 0302105;%%

\bibitem{CLLL}
M.~Cveti\v c, P.~Langacker, T.~Li and T.~Liu,
%``D6-brane splitting on type IIA orientifolds,''
hep-th/0407178.
%%CITATION = HEP-TH 0407178;%%

\bibitem{WITT}
E.~Witten,
%``An SU(2) Anomaly,''
Phys.\ Lett.\ B {\bf 117} (1982) 324.
%%CITATION = PHLTA,B117,324;%%
\bibitem{MSII}
F.~Marchesano and G.~Shiu,
%``Building MSSM flux vacua,''
hep-th/0409132.
%%CITATION = HEP-TH 0409132;%%

\bibitem{foot}
We also systematically studied examples whose Standard Models
arises from  D6-branes that are at angles with the gauge sector
$U(4)_C\times U(2)_L\times U(2)_R$ \cite{CLL}. We found no
examples with quantized fluxes.


\bibitem{CLL}
M.~Cveti\v c, T.~Li and T.~Liu,
%``Supersymmetric Pati-Salam models from intersecting D6-branes: A road to the
%standard model,''
hep-th/0403061.
%%CITATION = HEP-TH 0403061;%%

\bibitem{KST}
S.~Kachru, M.B.~Schulz and S.~Trivedi,
%``Moduli stabilization from fluxes in a simple IIB orientifold,''
JHEP {\bf 0310}, 007 (2003), hep-th/0201028.
%%CITATION = HEP-TH 0201028;%%

\bibitem{UR}
A.~M.~Uranga,
%``D-brane probes, RR tadpole cancellation and K-theory charge,''
Nucl.\ Phys.\ B {\bf 598}, 225 (2001), hep-th/0011048.
%%CITATION = HEP-TH 0011048;%%


\bibitem{CLW}
M.~Cveti\v c, P.~Langacker and J.~Wang,
%``Dynamical supersymmetry breaking in standard-like models with  intersecting
%D6-branes,''
Phys.\ Rev.\ D {\bf 68}, 046002 (2003), hep-th/0303208.
%%CITATION = HEP-TH 0303208;%%

\bibitem{Cvetic:2005bn}
M.~Cvet\v ic, T.~Li and T.~Liu,
%``Standard-like models as type IIB flux vacua,''
arXiv:hep-th/0501041.
%%CITATION = HEP-TH 0501041;%%


\bibitem{KKLT}
S.~Kachru, R.~Kallosh, A.~Linde and S.~P.~Trivedi,
%``De Sitter vacua in string theory,''
Phys.\ Rev.\ D {\bf 68}, 046005 (2003),
hep-th/0301240.
%%CITATION = HEP-TH 0301240;%%



\end{thebibliography}
\end{document}